\documentclass[11pt,twoside]{article}
\usepackage{asp2010}

\resetcounters

\begin{document}

\title{An analysis of metallic high ion absorption line profiles at DA
  white dwarfs with circumstellar material}
\author{Nathan~J.~Dickinson,$^1$ Martin~A.~Barstow,$^1$ and Barry~Y.~Welsh$^2$
\affil{$^1$Department of Physics and Astronomy, University of
  Leicester, Leicester, LE1 7RH, UK}
\affil{$^2$Space Sciences Laboratory, University of California,
  Berkeley, California, USA}}

\begin{abstract}

Some hot DA stars exhibit circumstellar absorption in the metal
resonance lines in their spectra. In many cases, these circumstellar
features are unresolved from those originating in the
photosphere. To better understand the effect this circumstellar
blending has on photospheric abundance estimates, we
present here an analysis of the unresolved metal high ion absorption
features of six hot white dwarfs. In all cases, given the strong
circumstellar C\,{\sc iv} detections, the photospheric C\,{\sc iv}
abundances are reduced; conversely the weak circumstellar Si\,{\sc iv}
leads to modest photospheric abundance revisions. A possible new technique for modelling these
line profiles is discussed, that can better reproduce the
observations and provide a greater insight into the conditions of the
circumstellar medium.

\end{abstract}

\section{Introduction}

The study of the circumstellar environment of white dwarf stars has
proven to be of great interest in recent years. In addition to the
circumstellar dust (e.g. \citealt{Zuckermanetal03}) and gaseous discs
(e.g. \citealt{Gaensickeetal06}) seen at cooler white dwarfs due to
tidally disrupted extra-solar minor planets and asteroids (e.g. \citealt{Jura03}), and
the possible Kuiper belt type discs seen at hot white dwarfs with
planetary nebulae (PNe; e.g. \citealt{Suetal07}), circumstellar high ion absorption can be seen in the
ultraviolet spectra of some hot DAs, thought to originate in either
interstellar medium (ISM) or ancient, diffuse planetary nebula remnants near enough to the hot stars
to be ionised \citep{Bannisteretal03,Dickinsonetal12circ}.
In many cases, the velocity of the
circumstellar material is sufficiently close to the photospheric
velocity for the absorption lines to be unresolved, resulting in asymmetric,
blended line profiles. Several previous studies of photospheric metal abundances (e.g. \citealt{Barstowetal03}) fit
these blended line profiles with only photospheric models, possibly
leading to over-estimates of stellar metal abundance. Indeed, the study of
\cite{Dickinsonetal12circ} found that the circumstellar C\,{\sc iv} near G191-B2B
accounts for 128.77 of the 160.01\,m\AA\ equivalent width of
the 1548\,\AA\
absorption feature; similarly, the equivalent width of the
circumstellar component of the 1550\,\AA\ line (with a total equivalent
width of 141.93\,m\AA) is 106.47\,m\AA. 

Some DA stars have metal abundances in excess of those seen in white dwarfs stars
with similar effective temperature (\textit{T}$_{\rm eff}$) and log \textit{g} values. While
accretion from circumstellar material may be enriching the
photospheres of some of these degenerates (such as GD\,394), it may
also be the case that in some of the stars circumstellar absorption at
a velocity indistinguishable from the photospheric velocity may be
contributing to the equivalent widths of the absorption features,
giving the impression of higher metal abundances.

Following this, inconsistencies in abundances derived using different
ionisation stages may also be due to the contamination of the line
profiles of resonance lines. The C abundance of G191-B2B
derived by \cite{Barstowetal03} using the C\,{\sc iv} resonance doublet, without accounting for the
circumstellar contamination (C/H\,=\,4.00$\times10^{-7}$),  is just over double that derived using the
C\,{\sc iii} features (C/H\,=\,1.99$\times10^{-7}$). Furthermore, using the C
abundance derived from the C\,{\sc iv} doublets in the spectra of
WD\,1942+499 and WD\,2257$-$073, \cite{Lallementetal11} predicted a
strong C\,{\sc iii} multiplet that was not observed, leading to the
conclusion that the observed C\,{\sc iv} was circumstellar.

Given the importance a
robust understanding of photospheric composition has when deriving
accurate stellar parameters, and the use abundance
inconsistencies may have as a diagnostic of the presence of
circumstellar material, the effect of accounting
for the circumstellar absorption on photospheric abundance is explored here.

\section{Method and Observations}

Of the 23 objects studied by \cite{Dickinsonetal12circ}, eight
displayed unambiguous circumstellar material, and at only two
objects (REJ\,0457$-$281 and REJ\,1738+665) was the absorption from
this material completely resolved from the
photospheric absorption lines, while the circumstellar material at
Feige\,24 was resolved at only one binary phase. The remaining six
objects (including Feige\,24) with blended
absorption features, their stellar parameters and observation
information, are given in table \ref{table:tefflogg}. The
\textit{Hubble Space Telescope} Space Telescope Imaging Spectrograph
(STIS) and Goddard High Resolution Spectrometer (GHRS) spectra were
fit using {\sc xspec} \citep{Arnaud96}, and the circumstellar
components (to C\,{\sc iv} at all stars, and Si\,{\sc iv} at REJ\,1614$-$085, WD\,2218+706 and Ton\,021) were modelled using the Gaussian absorption line model `{\sc
gabs},' provided with initial parameters calculated using the circumstellar line properties
(circumstellar velocity, \textit{b} value and column density) measured by
\cite{Dickinsonetal12circ}. The photospheric components were modelled
using {\sc tlusty} \citep{HubenyLanz95}, and were placed at the photospheric velocity
found by \cite{Dickinsonetal12circ}. Absorption lines from all
elements other than that considered in each spectral region (1545 to 1555\,\AA\ for C\,{\sc iv} and 1390 to 1405\,\AA\ for Si\,{\sc iv}) were
excluded from the fit to avoid the coupling of abundances across
chemical species. However, the other photospheric metals
examined by \cite{Barstowetal03} were
included in the `background' of the models at the abundances found therein. All parameters were allowed to
fit freely, except for stellar \textit{T}$_{\rm eff}$ and log\,\textit{g}, which
were constrained to the error range stated in table
\ref{table:tefflogg}. The two spectra of Feige\,24 were fit
separately, not co-added as in previous work, to allow a consistency
check of the abundances derived with and without the circumstellar
contamination. 1$\sigma$ errors were computed on all abundance estimates.

\begin{table}
\caption{The objects studied, ordered by \textit{T}$_{\rm
    eff}$, their stellar parameters (from \citealt{Barstowetal03}) and the observation information for the data
  used.}
\smallskip
\begin{center}
{\small
\begin{tabular}{l c c c c c c c}
\tableline
\noalign{\smallskip}
WD & \textit{T}$_{\rm eff}$ & log\,\textit{g} & Data Source    &
Resolving Power\\
\noalign{\smallskip}
\tableline
\noalign{\smallskip}
Ton\,021        & 69\,711$\pm$530\,K     & 7.47$\pm$0.05 &\textit{STIS} [E140M] & 40\,000\\ 
Feige\,24$^\dagger$   & 60\,487$\pm$1\,100\,K    & 7.50$\pm$0.06  & \textit{STIS} [E140M] & 40\,000\\
REJ\,0558$-$373 & 59\,508$\pm$2\,200\,K    & 7.70$\pm$0.09 & \textit{STIS} [E140M] & 40\,000\\
WD\,2218+706 & 58\,582$\pm$3\,600\,K    & 7.05$\pm$0.12 & \textit{STIS} [E140M] & 40\,000\\
G191-B2B        & 52\,500$\pm$900\,K       & 7.53$\pm$0.09 & \textit{STIS} [E140H] & 110\,000\\
REJ\,1614$-$085 & 38\,840$\pm$480\,K       & 7.92$\pm$0.07 & \textit{GHRS} [G160M] & 22\,000\\
\noalign{\smallskip}
\tableline       
\end{tabular}
}
\end{center}
\label{table:tefflogg}
\end{table}

\section{Results}

\begin{figure}
\begin{center}
\includegraphics[angle=90, height=0.4\textheight]{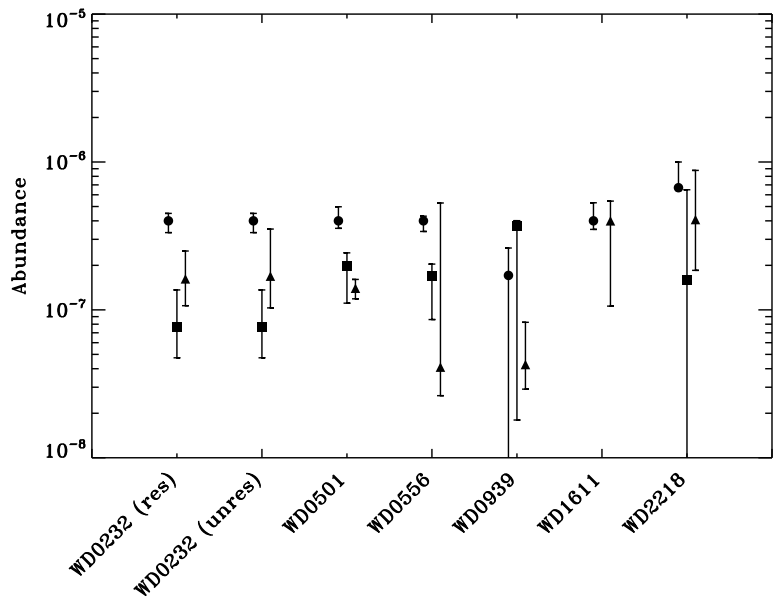}
  \caption{The C abundances measured in this study using the C\,{\sc iv}
    doublet (triangles), compared to the abundances derived using the  C\,{\sc iii} (squares) and C\,{\sc iv} (circles) absorption features by \cite{Barstowetal03}.}
  \label{fig:civ}
\end{center}
\end{figure}

\begin{figure}
\begin{center}
\includegraphics[angle=90, height=0.4\textheight]{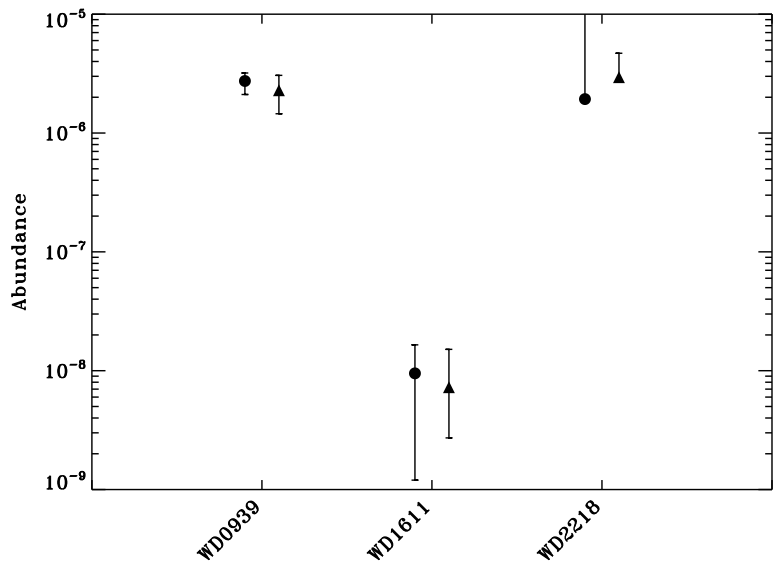}
  \caption{The Si abundances measured in this study (triangles), compared to the abundances derived by \cite{Barstowetal03} (circles).}
  \label{fig:siiv}
\end{center}
\end{figure}

\begin{figure}
\begin{center}
\includegraphics[angle=90,height=0.39\textheight]{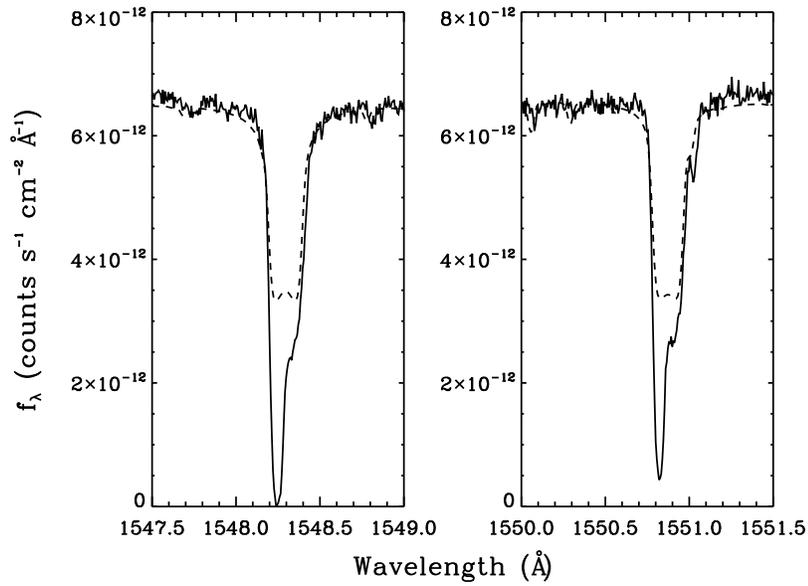}
 \caption{The  C {\sc iv} doublet of G191-B2B, fit with a model spectrum with the
   C abundance (C/H = 4.00x10$\rm^{-7}$) derived from the C\,{\sc iv} doublet by
   \cite{Barstowetal03}. The observed data is shown with a solid line,
   while the model is plotted with a dashed line.}
  \label{fig:photg191}
\end{center}
\end{figure}

\begin{figure}
\begin{center}
\includegraphics[angle=90, height=0.39\textheight]{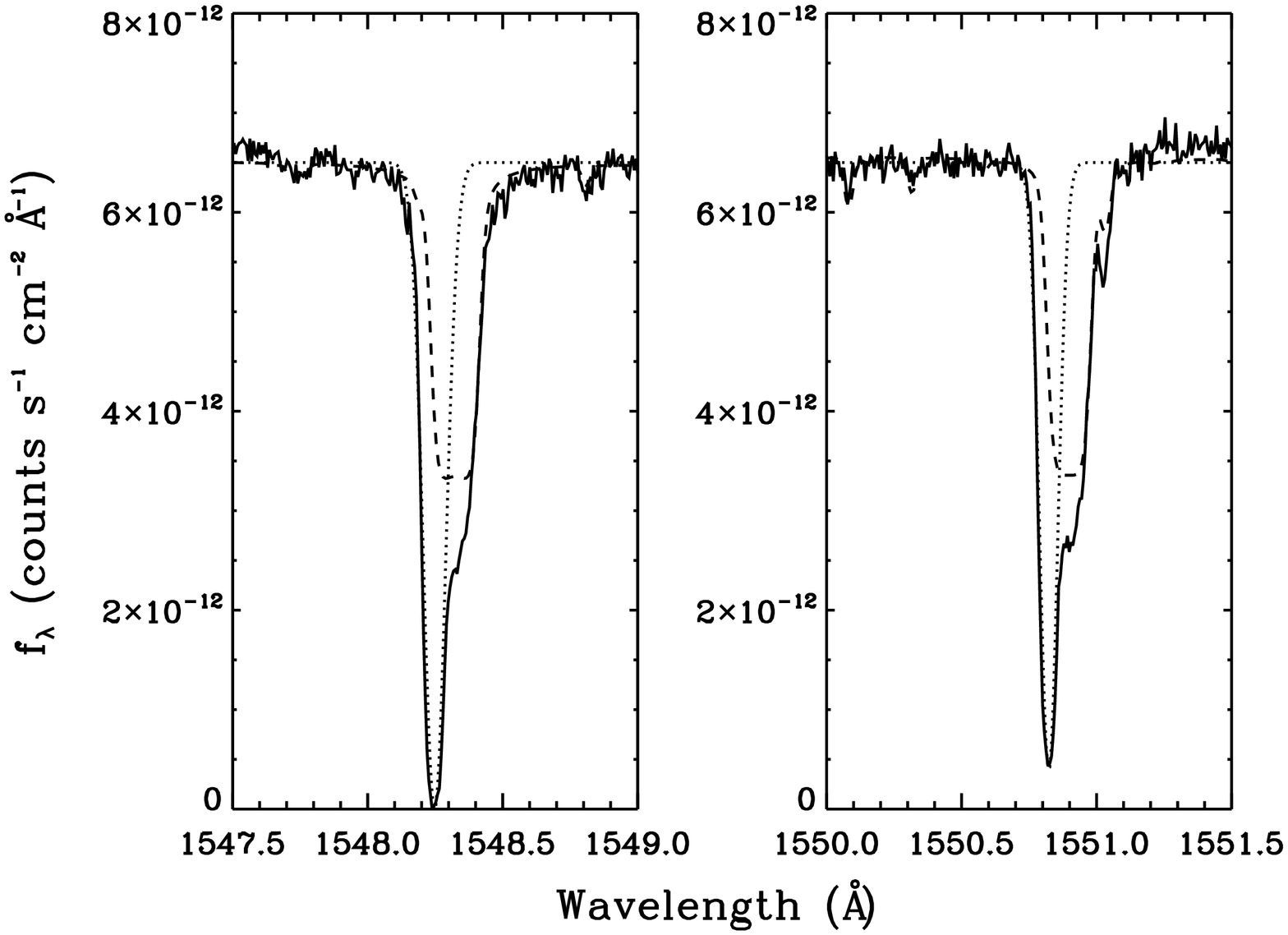}
 \caption{The best fitting model of the C {\sc iv} doublet of G191-B2B, with a photospheric C abundance of
 1.4x10$\rm^{-7}$ relative to hydrogen (dashed line). The dotted line represents the 
 circumstellar component and the solid line
 is the observed spectrum.}
  \label{fig:csg191}
\end{center}
\end{figure}

The abundances measured here are presented in table \ref{table:newabs}. The introduction of circumstellar components to the fits reduces the
measured C\,{\sc iv} abundances of the stars when compared to the
previous study of \cite{Barstowetal03} (figure \ref{fig:civ}). The weaker circumstellar
Si\,{\sc iv} makes a small difference to the measured photospheric
abundance (figure \ref{fig:siiv}). However, it must be noted that the combined effect of
modelling the circumstellar components with a simple Gaussian profile,
not a physically robust line profile prediction, and the
increased degrees of freedom to the model, make the statistical errors on the fits
large in some cases. The new C abundances for Feige\,24 agree across both
data sets. The improvement achieved progressing from a line profile model with only a
photospheric component (figure \ref{fig:photg191}), to one containing
both photospheric and circumstellar absorbers is show in figure \ref{fig:csg191}.

\begin{table}
\begin{center}
\caption{The abundances measured in this study, with 1$\sigma$ uncertainties. }
\begin{tabular}{l c c c c c c c c c}
\hline
WD                 & C\,{\sc iv}/H  & $+$1$\sigma$        &
$-$1$\sigma$& Si\,{\sc iv}/H &  $+$1$\sigma$       & $-1\sigma$\\
&[$\times10^{-7}$]&[$\times10^{-7}$]&[$\times10^{-7}$]&[$\times10^{-7}$]&[$\times10^{-7}$]&[$\times10^{-7}$]\\
\hline   
Ton\,021                & 0.43      & 0.40 & 0.14& 29.6     & 7.67 & 8.37\\
Feige\,24$^{\dagger}$  & 1.62       & 0.88& 0.55\\
Feige\,24$^{\ddagger}$  & 1.69       & 1.83 & 0.66\\
REJ\,0558$-$373            & 0.41       & 4.87 & 0.15\\
WD\,2218+706              & 4.08       & 4.96 & 2.23& 5.08       & 17.4 & 2.33\\
G191-B2B                & 1.40       & 0.21& 0.21\\
REJ\,1614$-$085            & 4.00       & 1.44 & 2.94& 0.07       & 0.08 & 0.05\\
\hline                 
\end{tabular}
\label{table:newabs}
\\\footnotesize{$^\dagger$resolved spectrum (0.74 binary phase), $^\ddagger$blended
spectrum (0.24 binary phase).}
\end{center}
\end{table}

\section{Conclusions}

The significant effect of circumstellar contamination to photospheric
abundance measurements, based on resonance line profile models, is evidenced
by the change in abundances shown here. Indeed, when accounting for
the circumstellar component in the C\,{\sc iv} doublet in the spectrum
of G191-B2B, the C abundance derived is in keeping with that found
using the C\,{\sc iii} features by \cite{Barstowetal03} and with that
found using \textit{FUSE} data (Barstow et al., ~these
  proceedings). Coupled with the use of inconsistencies in abundances
derived from different ionisation stages by \cite{Lallementetal11},
this study demonstrates the validity of using unusual photospheric
abundance measurements as a diagnostic of the presence of
circumstellar material in the absorption line profile. This will be
particularly useful in cases where circumstellar material is completely
unresolved from that in the photosphere, and symmetric line profiles
are present.

The C abundance seen
at REJ\,1614$-$085 here is larger than that of any other DA of similar
\textit{T}$_{\rm eff}$ (\citealt{Barstowetal03}), and is inconsistent with the abundance derived
from this star's \textit{FUSE} data (Barstow et al.,~these
  proceedings). Coupled with the difficulties in ascertaining the
distribution of photospheric nitrogen (\citealt*{Dickinsonetal12nv}) and the presence of Si
in this star at an abundance comparable to cooler DAs that are likely
to be accreting (\citealt{Barstowetal03}, \citealt*{Dupuisetal10}), it may be that
this star is accreting from an as yet undetected source
\citep{Burleighetal10,Burleighetal11}.

However, the larger errors on some of the measurements in this study are far
from ideal. One way in which this may be minimised is to use physically
robust, model line profiles for both the photospheric and
circumstellar components, providing a greater insight
into the conditions of the circumstellar medium beyond the
measurement of \textit{b} values and column densities. Indeed, the
comparison of line profiles from different physical models to those
observed may provide a powerful technique in ascertaining the source
of the circumstellar material seen in the spectra of these hot
DAs. 

\acknowledgements 
N.J.D. and M.A.B. acknowledge the support of STFC. B.Y.W. would like to
acknowledge Guaranteed Time Observer funding for this research through NASA
Goddard Space Flight Center grant 005118. N.J.D. wishes to thank Jay Holberg
and Ivan Hubeny for useful discussions.

\bibliography{dickinson}

\begin{thebibliography}{}
\expandafter\ifx\csname natexlab\endcsname\relax\def\natexlab#1{#1}\fi
\expandafter\ifx\csname url\endcsname\relax
  \def\url#1{\texttt{#1}}\fi
\expandafter\ifx\csname urlprefix\endcsname\relax\def\urlprefix{URL }\fi
\providecommand{\eprint}[2][]{\url{#2}}

\bibitem[{{Arnaud}(1996)}]{Arnaud96}
{Arnaud}, K. 1996, ASPC, 101, 17

\bibitem[{{Bannister} et~al.(2003){Bannister}, {Barstow}, {Holberg}, \&
  {Bruhweiler}}]{Bannisteretal03}
{Bannister}, N., {Barstow}, M., {Holberg}, J., \& {Bruhweiler}, F. 2003, MNRAS,
  341, 477

\bibitem[{{Barstow} et~al.(2003){Barstow}, {Good}, {Holberg}, {Hubeny},
  {Bannister}, {Bruhweiler}, {Burleigh}, \& {Napowotzki}}]{Barstowetal03}
{Barstow}, M., {Good}, S., {Holberg}, J., {Hubeny}, I., {Bannister}, N.,
  {Bruhweiler}, F., {Burleigh}, M., \& {Napowotzki}, R. 2003, MNRAS, 341, 870

\bibitem[{{Burleigh} et~al.(2010){Burleigh}, {Barstow}, {Farihi}, {Bannister},
  {Dickinson}, {Steele}, {Dobbie}, {Faedi}, \& {G\"{a}nsicke}}]{Burleighetal10}
{Burleigh}, M., {Barstow}, M., {Farihi}, J., {Bannister}, N., {Dickinson}, N.,
  {Steele}, P., {Dobbie}, P., {Faedi}, F., \& {G\"{a}nsicke}, B. 2010, in 17th
  European White Dwarf Workshop, edited by K.~{Werner}, \& T.~{Rauch} (New
  York: AIP), vol. 1273 of AIP Conf. Ser., 473

\bibitem[{{Burleigh} et~al.(2011){Burleigh}, {Barstow}, {Farihi}, {Bannister},
  {Dickinson}, {Steele}, {Dobbie}, {Faedi}, \& {G\"{a}nsicke}}]{Burleighetal11}
--- 2011, in Planetary Systems Beyond the Main Sequence, edited by S.~{Schuh},
  H.~{Dreschel}, \& U.~{Heber} (New York: AIP), vol. 1331 of AIP Conf. Ser.,
  289

\bibitem[{{Dickinson} et~al.(2012{\natexlab{a}}){Dickinson}, {Barstow}, \&
  {Hubeny}}]{Dickinsonetal12nv}
{Dickinson}, N., {Barstow}, M., \& {Hubeny}, I. 2012{\natexlab{a}}, MNRAS, 421,
  3222

\bibitem[{{Dickinson} et~al.(2012{\natexlab{b}}){Dickinson}, {Barstow},
  {Welsh}, {Burleigh}, {Farihi}, {Redfield}, \&
  {Unglaub}}]{Dickinsonetal12circ}
{Dickinson}, N., {Barstow}, M., {Welsh}, B., {Burleigh}, M., {Farihi}, J.,
  {Redfield}, S., \& {Unglaub}, K. 2012{\natexlab{b}}, MNRAS, 423, 1397

\bibitem[{{Dupuis} et~al.(2010){Dupuis}, {Chayer}, \&
  {Henault-Brunet}}]{Dupuisetal10}
{Dupuis}, J., {Chayer}, P., \& {Henault-Brunet}, V. 2010, in 17th European
  White Dwarf Workshop, edited by K.~{Werner}, \& T.~{Rauch} (New York: AIP),
  vol. 1273 of AIP Conf. Ser., 412

\bibitem[{{G\"{a}nsicke} et~al.(2006){G\"{a}nsicke}, {Marsh}, {Southworth}, \&
  {Rebassa-Mansergas}}]{Gaensickeetal06}
{G\"{a}nsicke}, B., {Marsh}, T., {Southworth}, J., \& {Rebassa-Mansergas}, A.
  2006, Science, 314, 1908

\bibitem[{{Hubeny} \& {Lanz}(1995)}]{HubenyLanz95}
{Hubeny}, I., \& {Lanz}, T. 1995, ApJ, 439, 875

\bibitem[{{Jura}(2003)}]{Jura03}
{Jura}, M. 2003, ApJL, 584, 91

\bibitem[{{Lallement} et~al.(2011){Lallement}, {Welsh}, {Barstow}, \&
  {Casewell}}]{Lallementetal11}
{Lallement}, R., {Welsh}, B., {Barstow}, M., \& {Casewell}, S. 2011, A\&A, 533,
  140

\bibitem[{{Su} et~al.(2007){Su}, {Chu}, {Rieke}, {Huggins}, {Gruendl},
  {Napiwotzki}, {Rauch}, {Latter}, \& {Volk}}]{Suetal07}
{Su}, K., {Chu}, Y.-H., {Rieke}, G., {Huggins}, P., {Gruendl}, R.,
  {Napiwotzki}, R., {Rauch}, T., {Latter}, W., \& {Volk}, K. 2007, ApJL, 657,
  41

\bibitem[{{Zuckerman} et~al.(2003){Zuckerman}, {Koester}, {Reid}, \&
  {H\"{u}nsch}}]{Zuckermanetal03}
{Zuckerman}, B., {Koester}, D., {Reid}, I., \& {H\"{u}nsch}, M. 2003, ApJ, 596,
  477

\end{thebibliography}

\end{document}